\documentclass[12pt]{article}
\usepackage{amsmath,amssymb,overcite}

\renewenvironment{thebibliography}[1]
 { \section*{References}
   \begin{list}{\arabic{enumi}.}
    {\usecounter{enumi} \setlength{\parsep}{0pt}
     \setlength{\itemsep}{3pt}
     \settowidth{\labelwidth}{#1.}
     \setlength{\leftmargin}{\labelwidth}
     \addtolength{\leftmargin}{3pt}
     \sloppy
    }}{\end{list}}

\def\ga{\gamma}
\def\de{\delta}
\def\ep{\epsilon}

\def\th{\theta}

\def\la{\lambda}

\def\si{\sigma}

\def\ta{\tau}

\def\vp{\varphi}

\def\om{\omega}
\def\Ga{\Gamma}
\def\De{\Delta}
\def\Th{\Theta}

\def\mn{{\mu\nu}}

\def\fr#1#2{{{#1} \over {#2}}}
\def\frac#1#2{\textstyle{{{#1} \over {#2}}}}

\def\prt{\partial}

\def\half{{\textstyle{1\over 2}}}
\def\lsim{\mathrel{\rlap{\lower4pt\hbox{\hskip1pt$\sim$}}
    \raise1pt\hbox{$<$}}}
\def\gsim{\mathrel{\rlap{\lower4pt\hbox{\hskip1pt$\sim$}}
    \raise1pt\hbox{$>$}}}

\def\etal {{\it et al.}}
\newcommand{\beq}{\begin{equation}}
\newcommand{\eeq}{\end{equation}}
\newcommand{\bea}{\begin{eqnarray}}
\newcommand{\eea}{\end{eqnarray}}
\newcommand{\bse}{\begin{subequations}}
\newcommand{\ese}{\end{subequations}}
\newcommand{\rf}[1]{(\ref{#1})}

\def\to{\rightarrow}

\def\mix{\leftrightarrow}
\def\tofrom{\rightleftarrows}
\def\nub{\bar\nu}
\def\vp{\vec p}
\def\cmat{{\mathcal C}}
\def\cH{{\mathcal H}}
\def\heff{h_{\rm eff}}

\def\mt{\widetilde m^2}

\def\ring#1{{\mathaccent'27 #1}}
\def\cri{\ring{c}}

\def\a3em{\check{a}}
\def\cee{\cri}
\def\Dtm{\De m^2_{\Th}}

\begin{document}

\title{Lorentz violation and neutrinos\footnote{
presented at the Third Meeting on CPT and Lorentz
Symmetry, Bloomington, Indiana, August, 2004.}}
\author{
  Matthew Mewes\\
  Department of Physics and Astronomy,
  Carleton College,\\
  One North College Street,
  Northfield, MN 55057
}
\date{}
\maketitle

\begin{abstract}
  Neutrino oscillations provide
  an opportunity for sensitive tests of Lorentz invariance.
  This talk reviews some aspects of
  Lorentz violation in neutrinos
  and the prospect of testing Lorentz invariance
  in neutrino-oscillation experiments.
  A general Lorentz-violating theory
  for neutrinos is discussed,
  and some signals of Lorentz violation are identified.
\end{abstract}

\section{Introduction}

Neutrinos offer a promising avenue
for the detection of new physics.
Evidence for neutrino oscillations
already indicates that the
minimal Standard Model (SM) of particle
physics needs modification.\cite{nureview}
The experiments providing this
evidence are in an excellent position
to detect tiny violations of 
Lorentz invariance\cite{nu2}
that may exist as the low-energy the remnants
of Planck-scale physics.\cite{ksp}
Here we discuss a general theoretical
framework describing the free propagation of
neutrinos in the presence of Lorentz violation.
We examine the effects of Lorentz violation
on neutrino oscillations and identify
unconventional behavior and experimental signals.

At attainable energies,
violations of Lorentz invariance
are described by a framework called the
Standard-Model Extension (SME).\cite{ck}
While the SME was originally motivated by
string theory,\cite{ksp}
it also encompasses other origins
for Lorentz violation
such as spacetime varying couplings.\cite{kpl}
The SME provides the basis for
a large number of experiments.\cite{cpt01}
In neutrinos, it gives a consistent
theoretical framework for the study
of Lorentz violation in
oscillations and other phenomena.
Neutrino-oscillation experiments
provide sensitivity to Lorentz-violating
effects\cite{nu2,nu1,nu3,fc,chki}
that rival the best tests in
any other sector of the SME.\cite{expt,kmph}

Remarkably, the current evidence for
neutrino oscillations lies at
levels where Planck-suppressed effects
might be expected to appear.
Furthermore, the possibility remains that
Lorentz violation may be responsible
at least in part for the observed oscillations.\cite{nu2,nu1,nu3}
Further analysis and experimentation is
needed to determine the extent to which
Lorentz violation may play a role in
neutrino oscillations.

\section{Framework}
In the SME,
the propagation of neutrinos is governed
by a modified multigeneration Dirac equation:\cite{nu2}
\beq
(i\Ga^\nu_{AB}\prt_\nu-M_{AB})\nu_B=0 \quad ,
\label{de}
\eeq
where three neutrino fields and
their charge conjugates
are included in order to allow for
general Dirac- and Majorana-type terms;
$\nu_A=\{\nu_e,\nu_\mu,\nu_\ta,\nu_e^C,\nu_\mu^C,\nu_\ta^C\}$.
Each of the quantities
$\Ga^\nu_{AB}$ and $M_{AB}$ are
$4\times 4$ constant matrices in spinor space.

Here we have included all
terms arising from operators of
renormalizable dimension,
but in general, higher derivative
terms can occur\cite{bef} and may
be important.\cite{kle}
It is also straightforward to include
additional generations in order
to accommodate sterile neutrinos.
Common Lorentz-conserving scenarios
exist as subsets of the general case.

The matrices $\Ga^\mu_{AB}$ and $M_{AB}$
can be decomposed using the basis of $\ga$ matrices.
Following standard conventions\cite{ck,kle},
we define 
\begin{align}
  \Ga^\nu_{AB} &:=
  \ga^\nu \de_{AB}
  + c^\mn_{AB}\ga_\mu
  + d^\mn_{AB}\ga_5\ga_\mu
  + e^\nu_{AB}
  + if^\nu_{AB}\ga_5
  + \half g^{\la\mn}_{AB}\si_{\la\mu}
  \quad ,
  \notag\\
  M_{AB} &:=
  m_{AB}
  +im_{5AB}\ga_5
  + a^\mu_{AB}\ga_\mu
  + b^\mu_{AB}\ga_5\ga_\mu
  + \half H^\mn_{AB}\si_\mn
  \quad .
  \label{GaM}
\end{align}
In these equations,
the masses $m$ and $m_5$ are Lorentz and $CPT$ conserving.
The coefficients $c$, $d$, $H$
are $CPT$ conserving but Lorentz violating,
while $a$, $b$, $e$, $f$, $g$
are both $CPT$ and Lorentz violating.
Requiring hermiticity of the
theory imposes the conditions 
$\Ga^\nu_{AB} = \ga^0(\Ga^\nu_{BA})^\dag\ga^0$
and $M_{AB} = \ga^0(M_{BA})^\dag\ga^0$,
which implies all coefficients are hermitian
in generation space.

Equation \rf{de} provides a basis
for a general Lorentz- and $CPT$-violating
relativistic quantum mechanics for freely
propagating neutrinos.
Construction of the relativistic hamiltonian
is complicated by the unconventional time-derivative term,
but this difficulty may be overcome in a manner
similar to that employed in the QED extension.\cite{kle}
The result is
\beq
\cH = \cH_0
-\half(\ga^0\de\Ga^0\cH_0+\cH_0\ga^0\de\Ga^0)
-\ga^0(i\de\Ga^j\prt_j-\de M)
\quad ,
\label{dH}
\eeq
where $\cH_0=-\ga^0(i\ga^j\prt_j-M_0)$
is the general Lorentz-conserving hamiltonian,
$M_0$ is the Lorentz-conserving part of $M$,
and $\de\Ga,\de M$ are the Lorentz-violating
parts of $\Ga,M$.

A general treatment is possible but
rather cumbersome and beyond the intended
scope of this work.
Therefore, we consider a simple
physically reasonable case where
oscillation between left- and right-handed
neutrinos is highly suppressed.
The resulting theory describes
oscillations between three flavors of
left-handed neutrinos due to
mass or coefficients Lorentz violation.
Within this restriction,
a calculation gives a $6\times6$
effective hamiltonian describing
the time evolution of
active neutrinos and antineutrinos
with momentum $\vp\,$:\cite{nu2}
\beq
\left(
  \begin{array}{c}
    \nu_a(t;\vp)\\
    \nub_a(t;\vp)
  \end{array}
\right)
=\exp(-i\heff t)_{ab}
\left(
  \begin{array}{c}
    \nu_b(0;\vp)\\
    \nub_b(0;\vp)
  \end{array}
\right)
\quad ,
\label{Ut}
\eeq
where $\nu_a$ and $\nub_a$ represent
active neutrino (negative helicity)
and antineutrino (positive helicity)
states, and indices $a,b$ range over $\{e,\mu,\ta\}$.
The effective hamiltonian is given by
\begin{align}
  &(\heff)_{ab}=|\vp|\de_{ab}+
  \frac{1}{2|\vp|}
  \left(\begin{array}{cc}
      (\mt)_{ab} & 0 \\
      0& (\mt)^*_{ab}
    \end{array}\right)\notag\\
  &+\hspace{-3pt}\frac{1}{|\vp|}\hspace{-3pt}
  \left(\begin{array}{cc}
      [(a_L)^\mu p_\mu-(c_L)^\mn p_\mu p_\nu]_{ab} &\hspace{-10pt}
      -i\sqrt{2} p_\mu (\ep_+)_\nu[(g^{\mn\si}p_\si-H^\mn)\cmat]_{ab} \\
      i\sqrt{2} p_\mu (\ep_+)^*_\nu[(g^{\mn\si}p_\si+H^\mn)\cmat]^*_{ab} &
      [-(a_L)^\mu p_\mu-(c_L)^\mn p_\mu p_\nu]^*_{ab}
    \end{array}\right).
  \label{heff}
\end{align}
This result assumes relativistic
neutrinos with momentum $|\vp|$
much larger than both mass and
Lorentz-violating contributions.
At leading order, the four momentum $p_\mu$ 
may be taken as $p_\mu=(|\vp|;-\vp)$,
and a suitable choice for $(\ep_+)^\nu$
is $(\ep_+)^\nu=\frac{1}{\sqrt{2}}(0;\hat\ep_1+i\hat\ep_2)$,
where $\hat\ep_1$, $\hat\ep_2$ are real
and $\{ \vp/|\vp|, \hat\ep_1, \hat\ep_2 \}$
form a right-handed orthonormal triad.

The above hamiltonian is consistent with
the standard seesaw mechanism,
where the right-handed Majorana
masses are much larger than
Dirac or left-handed Majorana masses.
However, the above equations
apply to any situation where
only left-handed neutrinos
are allowed to propagate or intermix.

Only the first term in Eq.\ \rf{heff}
arises from the minimal Standard Model.
The second term corresponds to
the usual massive-neutrino case 
without sterile neutrinos.
The leading-order Lorentz-violating contributions
are given by the last term.
Lorentz-violating $\nu\mix\nu$
mixing is controlled by the coefficient combinations 
$(c_L)^\mn_{ab}\equiv(c+d)^\mn_{ab}$ and
$(a_L)^\mu_{ab}\equiv(a+b)^\mu_{ab}$.
The remaining coefficients,
$(g^{\mn\si}\cmat)_{ab}$ and $(H^\mn\cmat)_{ab}$,
arise from gauge-violating Majorana-like couplings
and generate Lorentz-violating
$\nu\mix\nub$ mixing resulting
in lepton-number violations.
Note that some combinations of
coefficients are unobservable,
either because of symmetries or
because they can be removed through
field redefinitions \cite{ck,kle,cm,bek}.

Although this theory is observer independent
and therefore independent of choice of coordinates,
it is important to specify a frame for
reporting experimental results.
By convention this frame is taken
as a Sun-centered celestial equatorial
frame with coordinates $\{T,X,Y,Z\}$.\cite{kmph,spaceexpt}

\section{Features}\label{features}
A complete analysis of this construction
is hampered by its generality and lies
outside our present scope.
Two lines of attack have been initiated
in order to understand the theoretical
and experimental implications of
Lorentz violation.\cite{nu2}
The first involves the construction of
simple models that illustrate
the various unconventional features
and their potential to explain experimental data.
Some possibilities are considered in the next section.
An alternative strategy is to search for
`smoking-gun' signals that are
indicators of Lorentz violation.

The many coefficients for Lorentz violation
that appear in the effective hamiltonian \rf{heff}
introduce a plethora of new effects,
including unusual energy dependence,
dynamics dependent on the direction of propagation, 
and neutrino-antineutrino mixing.
Below we list six classes of model-independent
features that represent characteristic
signals of Lorentz violation in
neutrino-oscillation experiments.
A positive signal in any one of
these classes would suggest the
presence of Lorentz violation.

{\it Spectral anomalies.}
Each of the coefficients for Lorentz violation
introduces energy dependence differing
from the usual mass case.
In the conventional massive-neutrino case,
oscillations of neutrinos in the vacuum are
determined by the energy-independent mixing angles
$\th_{12}$, $\th_{13}$, $\th_{23}$,
phase $\de$, and mass-squared
differences $\de m$, $\De m$.
In this case, energy dependence
enters the oscillation probabilities
through the oscillation lengths
$L_0\propto E/\de m^2, E/\De m^2$.
In contrast, coefficients for Lorentz violation
can cause oscillation lengths that are either
constant or decrease linearly with energy.
For example, a simple model with only $c_L$
coefficients has much of the same structure
as the mass case except that it has oscillation lengths
$L_0\propto (E \de c_L)^{-1}, (E \De c_L)^{-1}$.
Combinations of coefficients with
different dimension can lead to very
complex energy dependence in both the
oscillation lengths and the mixing angles.
Detection of a vacuum oscillation length
that differs from the usual
$\propto E$ dependence or of energy dependence
in the vacuum mixing angles
would constitute a clear signal
of Lorentz violation.

{\it $L$--$E$ conflicts.}
This class of signal refers to
a set of null and positive measurements
that conflict in any scenarios based on
mass-squared differences.
In the usual case, baseline and
energy dependence enter through the ratio
$L/L_0\propto L/E$.
So experiments that measure the
same oscillation mode at similar
ranges in $L/E$ will have comparable
sensitivity to neutrino oscillations.
Because of the unusual energy dependence,
in Lorentz-violating scenarios this
may no longer be the case.
If oscillations are caused by coefficients
for Lorentz violation, it is possible that
experiments operating in the same region of
$L/E$ space could see drastically different
oscillation probabilities.
A measurement of this effect
would indicate physics beyond the
simple mass case and would constitute
a possible signal of Lorentz violation.

{\it Periodic variations.}
This signal indicates a violation
of rotation invariance and
would commonly manifest itself as
either sidereal or annual variations
in neutrino flux.
The appearance of $\vec p$ in the
effective hamiltonian \rf{heff}
implies that oscillations can
depend on the direction of the propagation.
In terrestrial experiments,
where both the detector and the source
are fixed relative to the Earth,
the direction of the neutrino
propagation changes as the Earth rotates.
This can lead to periodic variations
at the sidereal frequency
$\om_\oplus\simeq2\pi/$(23 h 56 min).
For solar neutrinos,
the variation in propagation of the
detected neutrinos is due
to the orbital motion of the Earth
and can cause annual variations.

{\it Compass asymmetries.}
This class includes time-independent
effects of rotation-invariance violations.
They consist of unexplained directional
asymmetries in the observed neutrino flux.
For terrestrial experiments,
averaging over time eliminates any sidereal variations,
but may leave a dependence on the direction
of propagation as seen from the laboratory.
This can result in asymmetries between
the compass directions north, south,
east, and west.

{\it  Neutrino-antineutrino mixing.}
This class includes any measurement 
that can be traced to $\nu\mix\nub$ oscillations.
This would indicate lepton-number violation
that could be due to $g$ and $H$ coefficients.
All of these coefficients introduce
rotation violation, so this signal may be 
accompanied by direction-dependent signals.

{\it Classic $CPT$ test.}
This is the traditional test of $CPT$
involving searches for violations
of the relationship
$P_{\nu_b\to\nu_a}(t)=P_{\nub_a\to\nub_b}(t)$.
This equation holds provided $CPT$ is unbroken.
An additional result holds in the event of
lepton-number violation:
$P_{\nu_b\tofrom\nub_a}(t)=P_{\nu_a\tofrom\nub_b}(t)$,
if $CPT$ is unbroken.
A measurement that contradicts
either of these relations is a
signal of $CPT$ violation
and would therefore imply
Lorentz violation.

\section{Illustrative models}

In this section,
we discuss some simple subsets of the
general case \rf{heff} that exhibit some
of the unconventional effects.
While in most cases
these models are not expected to agree with
all existing data,
they do provide useful insight into
the novel behavior that Lorentz violation can introduce.
An interesting open challenge is to identify
general classes of realistic models
that could be compared to experiment.
The bicycle model\cite{nu1} and
its variants offer possibilities that
have no mass-squared differences and
few degrees of freedom.

\subsection{Fried-chicken models}
One simple class of models are those
dubbed `fried-chicken' (FC) models.
The idea behind these is to restrict
attention to direction-independent
behavior by only considering isotropic coefficients.
This restriction reduces the
effective hamiltonian to
\begin{align}
  (\heff)^{\rm FC}_{ab}&={\rm diag}
  \big[\big(
  \mt/(2E)
  +(a_L)^T
  -\frac43(c_L)^{TT}E
  \big)_{ab}\ ,
  \notag \\
  &\qquad\qquad\qquad\big(
  \mt/(2E)
  -(a_L)^T
  -\frac43(c_L)^{TT}E
  \big)^*_{ab}\big] 
  \quad .
  \label{rim}
\end{align}
A majority of the Lorentz-violating
models considered in the literature
are subsets of this general FC model.\cite{fc}

The differences in energy dependence
between the various types of
coefficients and mass
is apparent in Eq.\ \rf{rim}.
FC models provide a workable
context for studying the unconventional
energy dependence
without the complication of
direction-dependent effects.
However, it should be noted
that Eq.\ \rf{rim} is a highly frame dependent.
Isotropy in a given frame necessarily
implies anisotropy in other frames
boosted with respect to the isotropic one.
While it may be appealing to impose
isotropy in a frame such as the
cosmic-microwave-background frame,
it is difficult to motivate theoretically.

\subsection{Vector models}
In contrast to FC models,
vector models are designed to
study the effects of rotation-symmetry violation.
These models contain coefficients
that can be viewed as three-dimensional
vectors that point in given directions.
They are particularly useful
in determining the types of signals
that a given experiment might expect to see
if rotation symmetry is violated.

As an example consider a model where only
the coefficients
$(a_L)_{\mu\ta}^{X}$,
$(a_L)_{\mu\ta}^{Y}$,
$(c_L)_{\mu\ta}^{TX}$, and
$(c_L)_{\mu\ta}^{TY}$
are nonzero.
Each of these can be viewed as vectors
lying in the Earth's equatorial plane.
They are chosen to illustrate the periodic signals
discussed in the previous section.
With the above choice, we would see
maximal mixing between
$\nu_{\mu}\mix\nu_{\ta}$
and $\nub_{\mu}\mix\nub_{\ta}$,
which are relevant oscillation modes
for atmospheric neutrinos.
So, this simple special case may serve as a
test model for searches for sidereal
variations in atmospheric neutrinos.

The vacuum oscillation probability
for a terrestrial experiment is
\begin{align}
  P_{\nu_\mu\leftrightarrow\nu_\tau}&=\sin^2
  L\Big((A_s)_{\mu\tau}\sin\omega_\oplus T_\oplus
  +(A_c)_{\mu\tau}\cos\omega_\oplus T_\oplus\Big)
  \quad ,\\
  \intertext{where}
  (A_s)_{\mu\tau}&=
  \hat N^Y\Big((a_L)^X_{\mu\ta}-2E(c_L)^{TX}_{\mu\ta}\Big)
  -\hat N^X\Big((a_L)^Y_{\mu\ta}-2E(c_L)^{TY}_{\mu\ta}\Big)
  \quad ,\\
  (A_c)_{\mu\tau}&=
  -\hat N^X\Big((a_L)^X_{\mu\ta}-2E(c_L)^{TX}_{\mu\ta}\Big)
  -\hat N^Y\Big((a_L)^Y_{\mu\ta}-2E(c_L)^{TY}_{\mu\ta}\Big)
  \ .
\end{align}
Here $N^X$ and $N^Y$ are factors
that are determined by the direction
of the neutrino propagation as seen
in the laboratory.
In this example, both the unusual energy
dependence and the sidereal variations
are readily apparent.
The dependence on beam direction
through $N^X$ and $N^Y$
implies that a time average in
this model also gives rise
to compass asymmetries.
These could be sought in
atmospheric experiments and
other experiments where neutrinos
originate from different compass
directions.

\subsection{The bicycle model}
One class of interesting special
cases are those that involve a
Lorentz-violating seesaw mechanism.
The resulting dynamics can be
dramatically different than
what is naively expected from
the effective hamiltonian \rf{heff}.
One such model is the bicycle model.\cite{nu1}
This model is also interesting
because it crudely matches
the basic features seen in 
solar and atmospheric neutrinos
using only two degrees of freedom.

The bicycle model consists of
an isotropic $c_L$ with nonzero element 
$\fr43(c_L)^{TT}_{ee} \equiv 2\cee >0$
and an anisotropic $a_L$ with degenerate
nonzero real elements 
$(a_L)^Z_{e\mu}=(a_L)^Z_{e\ta}\equiv\a3em/\sqrt{2}$.
The vacuum oscillation probabilities are 
\begin{align}
  P_{\nu_e\to\nu_e}&=
  1-4\sin^2\theta\cos^2\theta\sin^2(\De_{31}L/2)
  \quad ,
  \nonumber\allowdisplaybreaks\\
  P_{\nu_e\mix\nu_\mu}
  &= P_{\nu_e\mix\nu_\ta} =2\sin^2\theta\cos^2\theta\sin^2(\De_{31}L/2)
  \quad ,
  \nonumber\allowdisplaybreaks\\
  P_{\nu_\mu\to\nu_\mu}
  &= P_{\nu_\ta\to\nu_\ta}
  =1-\sin^2\theta\sin^2(\De_{21}L/2) 
  \nonumber\\ 
  &\qquad\quad
  -\sin^2\theta\cos^2\theta\sin^2(\De_{31}L/2) 
  -\cos^2\theta\sin^2(\De_{32}L/2)
  \quad ,
  \nonumber\allowdisplaybreaks\\
  P_{\nu_\mu\mix\nu_\ta}&=
  \sin^2\theta\sin^2(\De_{21}L/2) 
  \nonumber\\
  &\qquad\quad
  -\sin^2\theta\cos^2\theta\sin^2(\De_{31}L/2)
  +\cos^2\theta\sin^2(\De_{32}L/2)
  \quad ,
\end{align}
where
\begin{align}
  \De_{21}&=\sqrt{(\cee E)^2+(\a3em\cos\Th)^2}+\cee E
  \quad , 
  \nonumber \\
  \De_{31}&=2\sqrt{(\cee E)^2+(\a3em\cos\Th)^2}
  \quad ,
  \nonumber \\
  \De_{32}&=\sqrt{(\cee E)^2+(\a3em\cos\Th)^2}-\cee E
  \quad ,
  \nonumber \\
  \sin^2\theta &=\half [1-{\cee E}/
  {\sqrt{(\cee E)^2+(\a3em\cos\Th)^2}}]
  \quad ,
  \label{sinth}
\end{align}
and where $\Th$ 
is defined as the angle between the celestial
north pole and the direction of propagation.
These probabilities also hold for antineutrinos,
which implies that it is possible
to violate $CPT$ and not produce the last signal
discussed in Sec.\ \ref{features}.

An important feature of this model is that
at high energies, $E\gg|\a3em |/\cee$,
a seesaw mechanism takes effect
and oscillations reduce to two-generation
mixing with
$P_{\nu_\mu\mix\nu_\ta}\simeq\sin^2(\De_{32}L/2)$,
$\De_{32}\simeq\a3em^2\cos^2\Th/2\cee E$. 
The energy dependence in this regime mimics
exactly that of the usual mass case.
However, the quantity that takes the
place of mass, the pseudomass
$\Dtm=\a3em^2\cos^2\Th/\cee$,
is dependent on the direction of propagation.
So it is possible to construct models
with conventional energy dependence
but unconventional direction dependence.

\section{Short baseline experiments}
Some circumstances are
amenable to more general analyses.
One case where this is true is when
the baseline of an experiment is
short compared to the oscillation
lengths given by the hamiltonian \rf{heff}.\cite{nu3}
In this situation, the transition
amplitudes can be linearized,
which results in leading order
probabilities given by
\beq
P_{\nu_b\to\nu_a}\simeq
\left\{
  \begin{array}{lr}
    1-\sum_{c, c\ne a}
    P_{\nu_a\to\nu_c}, \quad & a=b\ ,\\[10pt]
    |(\heff)_{ab}|^2L^2/(\hbar c)^2, & a\ne b\ .
  \end{array}
  \right.
  \label{P}
\eeq
This approximation allows direct
access to the coefficients
for Lorentz violation without
the complication of diagonalizing
the hamiltonian.
This makes an analysis of the general
hamiltonian \rf{heff} more practical.

This type of analysis may be
relevant for the LSND experiment,
which is consistent with
a small oscillation probability
$P_{\nub_\mu\to\nub_e}\simeq 0.26$
over a short baseline of about 30 m.\cite{lsnd}
This result is of particular interest
because it does not seem to fit into
the simple three-generation solution
to solar and atmospheric data.
The possibility exists that Lorentz violation
may provide a solution.

\end{document}